# The fine structure constant: a review of measurement results and possible space-time variations

**Kirill A. Bronnikov[1], Vladimir D. Ivashchuk[2], Viacheslav V. Khruschov[3]**

[1, 2, 3] Research Center for Applied Metrology – Rostest, Moscow, Russia

[1]kb20@yandex.ru, [2]vladimirdi@rostest.ru, [3]vyacheslavvk@rostest.ru

**Abstract.** A brief description of the main methods for determining the fine structure constant is given. It is shown that the exact value of the fine structure constant is important for the new International System of Units (SI) and for fundamental metrology. Recent measurement results and theoretical calculations of the fine structure constant, as well as its possible space-time variations, are presented. The results of laboratory experiments on the search for long-term variations of the fine structure constant are described. The astrophysical and cosmological observational data on possible variability of the fine structure constant are displayed. The possibility of slightly lower values of the fine structure constant in the remote past as compared to its modern value, as well as the existence of unresolved problems related to possible space-time variations of the fine structure constant and the spread of the results of its precise laboratory measurements, are mentioned. Despite the absence of experimentally confirmed long-term variations of the fine structure constant at a high level of accuracy, possible practical applications of the results are noted, namely, the construction of an optical frequency standard with high stability and frequency reproduction accuracy based on the ytterbium-171 ion and a laser frequency synthesizer which may replace the caesium frequency standard.

**Introduction**

In November 2018, at the XXVIth General Conference on Weights and Measures [1], the transition to new definitions of a number of basic units of the International System of Units (SI) was declared. The new version of the SI became effective on May 20, 2019. The new definitions of SI are based on precise values of fundamental physical constants (FFCs) [1-3]. Fixing the numerical values of Planck's constants $h$, Avogadro's constant $N_A$, the electron charge $e$, and Boltzmann's constant $k$ made it possible to redefine the units of mass, amount of matter, current, and temperature.

The SI system reform is aimed at improving the accuracy, stability and reproducibility of measurement results (anywhere and at any time). At the same time, the units of quantities determined using the FFC, in principle, become independent of many natural and social factors. The reform was preceded by the well-known transitions to the atomic standard of the unit of time (frequency) and the relativistic standard of the unit of length, which significantly improved the accuracy of implementation of these units.

The purpose of this paper is to review the current results related to the fine structure constant α, a dimensionless fundamental physical constant that plays an important role not only in quantum physics and chemistry, but also in fundamental and applied metrology. The main methods of measuring the fine structure constant are described, and the results of recent measurements in laboratory conditions are presented. The role of the fine structure constant in the new SI is discussed. The results of a search for long-term α variations in laboratory conditions are presented, as well as astrophysical and cosmological data on the possible variability of the fine structure constant.

It is known that the search for space-time variations of the fine structure constant is simultaneously a test of the Einstein equivalence principle. This principle, in particular, states that the result of any non-gravitational experiment does not depend on its position in time and space [4]. In theories that go beyond the Standard Model of Fundamental Interactions (SM), space-time variations of FFCs are not excluded [5]. Moreover, astronomical observations have provided some indications of the possibility of spatial variations of the fine structure constant [6]. At the same time, there are no signs of space-time variations of α in laboratory conditions [7]. The results of the search for space-time variations of FFC will determine both the necessary generalization of the SM and further improvement of SI, see, for example, [8].

**Basic methods for determining the fine structure constant**

The fine structure constant α, introduced in 1916 by A. Sommerfeld [9], is a dimensionless quantity that characterizes the force of the electromagnetic interaction and is determined by the following relation in SI:

$$\alpha = e^2 / (4\pi\varepsilon_0 \hbar c),$$

where $\varepsilon_0$ is the electric constant; $\hbar = h/(2\pi)$ is the reduced Planck constant, and $c$ is the speed of light.

The fine structure constant α is included in the formulas of the ionization energy of the hydrogen atom, fine and hyperfine structures of atomic energy levels. The value α is also a decomposition parameter in calculations based on perturbation theory in quantum electrodynamics. The value of α is also a decomposition parameter in calculations based on

3
perturbation theory in quantum electrodynamics. Measurements of α in various fields of physics can serve as a test for the consistency of quantum electrodynamics and its generalization in the SM.

Currently, there are two methods for obtaining sufficiently accurate values of α. In the first (theoretical and experimental) method [10-14], α is determined as a result of a combination of measurements of the anomaly parameter of the magnetic moment of the electron (the electron anomaly) $a_e$ and laborious calculations of $a_e$ in quantum electrodynamics [15-17], or, more precisely, in the framework of the SM of elementary particles. As is known, the magnetic moment of an electron is proportional to the product of its spin, the Bohr magneton, and the *g*-factor, which is written in a vector form as

$$\boldsymbol{\mu} = -g\mu_B \mathbf{S}/\hbar.$$

The anomaly of the magnetic moment of an electron is by definition

$$a_e = (g-2)/2.$$

In quantum electrodynamics, the excess of the *g*-factor over its Dirac value is explained by the fact that the magnetic moment of an electron increases due to the generation of virtual particles and vacuum polarization. In this approach, the *g*-factor and anomalies of the magnetic moment of the $a_e$ are experimentally measured. The results obtained are compared with the results of theoretical calculations in the framework of quantum electrodynamics (or SM) in the methodology of perturbation theory by α. By equating the obtained experimental $a_{e\,\text{exp}}$ and theoretical $a_{e\,\text{th}}$ values of $a_e$, we arrive at the equation $a_{e\,\text{exp}} = a_{e\,\text{th}}(\alpha)$ from which we find the value of α.

Currently, the *g*-factor of an electron is determined experimentally mainly using a single-electron cyclotron [11]. This device was created by G. Gabrielse and S. Peil in 1999 and has been repeatedly improved since then. It is a small conducting cavity in which a single electron is trapped using alternating electromagnetic fields (in fact, this is a modification of the Penning trap). During measurements, a magnetic field is activated, directed along the axis of the device. In the presence of this field, the electron moves in a spiral with a cyclotron frequency $f_c$ and simultaneously precesses around the field vector with a different frequency $f_s$ [11]. At that, the *g*-factor exceeds the value 2 by a small amount $(f_s - f_c)/f_c$, which is determined experimentally. In the experiment, it is required to calculate the geometry of the inner cavity of the trap very accurately and to ensure that the photonic "gas" (radiation) in the trap is cooled to very low temperatures (< 0.1 K), so that the electron orbits are stable during a multi-hour experiment under the conditions of relatively weak and slowly changing electromagnetic fields. One also takes into account relativistic corrections (rather small due to the low kinetic energy of the electron).

In the second (experimental) approach, the fine structure constant α is obtained as a result of measurement of $h/m_X$ using the ratio [18-24]:

$$\alpha^2 = (2R_\infty/c)(m_X/m_e)(h/m_X) \qquad (1)$$

where X denotes a specific atom, usually an atom of rubidium $^{87}$Rb [20-22, 24] or caesium $^{133}$Cs [18, 19, 23]: $R_\infty$ is the Rydberg constant; $m_X/m_e$ is the atom/electron mass ratio, respectively.

In Eq. (1), the Rydberg constant is known with relative standard uncertainty $1,9 \cdot 10^{-12}$ [25], and the mass ratio $m_X/m_e$ is known with relative standard uncertainty (RSU) σ of the order $1,5 \cdot 10^{-10}$ and $1,3 \cdot 10^{-10}$ for $^{87}$Rb and $^{133}$Cs, respectively [26]. (The values of the RSU were obtained according to CODATA-2018 [25] data on the mass of an electron in atomic units). The limiting factor in the second approach is the ratio $h/m_X$, which is determined by measuring the recoil velocity $v_r = \hbar K/m_X$ of an atom X absorbing a photon with momentum $\hbar K$, where $K$ is the modulus of the wave vector. The experimental principle for the second approach was previously described (for X = $^{87}$Rb) in [20] and is based on the use of the Ramsey-Borde atomic interferometer [19] and Bloch oscillations. The idea of the experiment is to implement multiple and coherent pulse transmissions by a stationary atom and measure the distribution of recoil velocities. As a result, the ratio $h/m_X$ is determined, using which as well as $R_\infty$ and $m_X/m_e$, the relation $\alpha = \alpha(X)$ can be found.

It should be noted that with the adoption of the new definition of the unit of mass (kilogram) in the new SI by fixing the numerical value of Planck's constant, precision determination of the ratio $h/m_X$ is actually equivalent to finding with high accuracy the mass value of the atom $m_X$ in units of the new kilogram.

**Modern results of measurements of the fine structure constant**

The current value of the fine structure constant, adopted in the CODATA-2018 agreement [25] is

$$\alpha(\text{CODATA} - 2018) = 7,2973525693(11) \cdot 10^{-3}.$$

This value can be further written in terms of the inverse value,

$$\alpha^{-1}(\text{CODATA}-2018) = 137{,}035999084(21),$$

which is obtained with a RSU of $1{,}5 \cdot 10^{-10}$ and corresponds to an anomalous magnetic moment $a_e = 1{,}15965218128(18) \cdot 10^{-3}$ with a RSU of $1{,}5 \cdot 10^{-10}$ [25].

Formally, the value of α adopted in the CODATA-2018 agreement, coincides with the value obtained in 2007 by G. Gabrielse and colleagues [13]. At the same time, the accuracy in matching CODATA-2018 has increased by a factor of 2.5 as compared to the 2007 result of the Harvard group [8]. The result of [13] is based on the experimental measurement of the *g*-factor using a single-electron cyclotron, as well as on the calculation of the electron anomaly $a_e$ in the fourth order of the α perturbation theory performed by T. Kinoshita and colleagues [16].

The CODATA-2018 agreement is based mainly on the results of two approaches: the theoretical and experimental one [13, 17] and the experimental one [22]. In 2018, in the work of the Berkeley group [22], as part of experiments with the caesium $^{133}$Cs atom, an exclusively experimental result, which was a record at that time in terms of accuracy, was obtained:

$$\alpha^{-1}(\text{Cs,Berkeley}-2018) = 137{,}035999046(27),$$

with RSU of $2 \cdot 10^{-10}$.

In the theoretical work of T. Kinoshita (USA), T. Aoyama and M. Nio (Japan) [17], the electron anomaly $a_e$ was calculated using a higher (fifth) order of perturbation theory as compared to the calculations performed in previous works [16], where the fourth order of perturbation theory was considered.

The value of $a_e$ can be expressed as a sum of six terms, five of which contain the factor α/π in different powers with coefficients depending on the contributions of the muon and the tau-lepton, while the sixth term corresponds to the weak and strong interactions:

$$\begin{aligned} a_e &= A_1 \alpha/\pi + A_2 (\alpha/\pi)^2 + A_3 (\alpha/\pi)^3 + A_4 (\alpha/\pi)^4 + \\ &+ A_5 (\alpha/\pi)^5 + a(\text{weak, hadron}). \end{aligned} \quad (2)$$

In (2), the dimensionless coefficients $A_i$ arise when calculating renormalized integrals for (Feynman) diagrams of quantum electrodynamics, taking into account the contributions of leptons of three generations. For $i > 1$ these coefficients depend on the mass ratios of the leptons:

$$A_i = A_i(m_e/m_\mu, m_e/m_\tau).$$

The mass ratios $m_e/m_\mu$, $m_e/m_\tau$ are usually taken from the data of the recent (at the time of calculations) CODATA agreement, and the accuracy of their determination is sufficient for obtaining precision values of α in this approach [27, 28]. The first coefficient obtained by Schwinger in 1948 [29] does not depend on the mass ratios of leptons. The coefficient $A_1$ corresponds to one diagram, and the coefficients $A_2$, $A_3$, $A_4$, $A_5$ correspond to 7, 72, 891, and 12672 diagrams, respectively.

In 2020, the work of the French group [23] published an (intriguing) result of measuring the ratio of Planck's constant to the mass of the $^{87}$Rb atom: $h/m_{Rb}$ (proportional to the inverse mass of the rubidium isotope in kilograms) and obtained the value

$$\alpha^{-1}(\text{Rb,Paris}-2020) = 137,035999206(11)$$

with RSU $0,81 \cdot 10^{-10}$, that improved the previous result of the Paris group [20] in terms of accuracy by about a factor of 8. Moreover, the accuracy of the claimed result is about twice as high as the accuracy of the determination of α in CODATA-2018. However, this result does not agree well with the previously accepted result of CODATA-2018 [25], even the confidence intervals at the 3σ level do not overlap. The resulting value shifts the previous α value by approximately $9 \cdot 10^{-9} \alpha$.

In 2023, the Harvard group [14] obtained a new value of the electron's *g*-factor, which improved the accuracy of the previous result of 14 years ago by a factor of 2. Using the results of quantum calculations of the *g*-factor and anomalies of the magnetic moment of the electron, obtained in the works of T. Kinoshita and colleagues [17], the authors of the article [14] obtained the following value of the fine structure constant:

$$\alpha^{-1}(\text{Harward-2023}) = 137,035999166(15)$$

with RSU $\sigma = 1,1 \cdot 10^{-10}$. The α value found in [14] is closer to the 2020 result of the Paris group [23] than to the previous result [13] obtained in 2007. Thus, at present there is a problem of matching the obtained most accurate results of the fine structure constant.

**The fine structure constant in the new SI**

The quantity α plays an important role in metrology [3]. In the new SI, it defines the electrical constant $\varepsilon_0$ (and, as a result, the magnetic constant $\mu_0 = \left(c^2 \varepsilon_0\right)^{-1}$), which should be known with the same RSU as α, i.e. $1,5 \cdot 10^{-10}$ in the CODATA-2018 agreement [25]. The value of the fine structure constant is also included in the expression for the Planck molar constant, the product of the Planck and Avogadro constants:

$$N_A h = A_r(e) M_u c \alpha^2 / (2 R_\infty),$$

where $M_u = N_A m_u = 0,99999999965(30)$ g/mol is the mass molar constant [25]; $A_r(e)$ is the relative atomic mass of an electron, known with RSU $2,9 \cdot 10^{-11}$ [25].

As mentioned above, the Rydberg constant is known with RSU = $1,9 \cdot 10^{-12}$ [25]. In the new SI, Planck's molar constant is precisely defined as

$$N_A h = 3,9903127128934314 \; 10^{-10} \; \text{J s mol}^{-1},$$

and, as a result, the mass molar constant $M_u$ has twice as much RSU as α, i.e., the RSU of $M_u$ is $3 \cdot 10^{-10}$ [25]. The correction factor k also implicitly depends on the fine structure constant, which occurs when redefining a mole in the new SI and when determining the mass value of an atom of $^{12}$C,

$$1 + k = M_u / M_u^{old} = N_A m(^{12}C) / (12 M_u^{old}),$$

where $M_u^{old} = 1$ g/mol is the molar mass constant in the old SI. When matching the values of CODATA-2018, the correction factor is determined by the relation $1 + k = 1 - 0,35(30) \cdot 10^{-9}$.

**A search for long-term alpha variations in laboratory conditions**

As noted in [30], when searching for space-time variations of the SM constants, the main interest is the experimentally determined values of the logarithmic derivatives of the coupling constants of fundamental interactions and the ratios of particle masses to the electron mass in spatial and temporal variables. Variations of these quantities on large scales of space and time can also influence the course of the thermal history of the universe [31] and lead to observable effects.

When searching for long-term variations in the coupling constants of electromagnetic, strong, and weak interactions, the largest number of experimental constraints were obtained for the coupling constants of electromagnetic interactions $\alpha = e^2 / (4\pi\varepsilon_0 \hbar c)$. For example, laboratory upper bounds on $\dot{\alpha}/\alpha$ were obtained by comparing optical clocks: $|\dot{\alpha}/\alpha| < 2 \cdot 10^{-17}$ year$^{-1}$ [32, 33]. Astrophysical observations of extragalactic sources give almost the same limit [34].

Let us consider in more detail the recently obtained restriction on the long-term variation of α [35]. In [35], a search was carried out for variations in α and the ratio of proton to electron masses, $k_p = m_p / m_e$, by comparing optical atomic clocks with a single ytterbium-171 ion and clocks with a cesium fountain for a period of more than 4 years. Currently, as is well known, the primary reproduction of a second occurs using the atomic frequency standards based on digital fountains. A special feature of the ytterbium-171 ion is the strongly different electronic structures of excited states, from which transitions to the ground state occur using the electric multipole transitions E2, E3. The transition energy using E3 includes significant relativistic contributions and is sensitive to α variations [36]. The ability to study transitions in the same ion with different sensitivity to variations $\alpha(E2, E3)$ has significantly reduced the complexity of the experimental setup. The clock systems used made it possible to reduce the relative uncertainty of the E3 transition energy to the level of $3 \cdot 10^{-8}$. One clock system was configured for the E2 transition, the other for E3. The purpose of the work [35] was to study the frequency ratio $\nu_{E3}/\nu_{E2}$ over a period of more than 4 years. As a result, an upper limit on the relative temporal variation of the fine structure constant was obtained, equal to a $1,0(1,1) \cdot 10^{-18}$ year$^{-1}$. The result obtained is compatible with zero within one standard deviation of $1,1 \cdot 10^{-18}$ per year, which allows us to speak at this level of accuracy about the constancy of the fine structure constant. To date, a stronger restriction on the variation of the fine structure constant, equal to $1,8(2,5) \cdot 10^{-19}$ year$^{-1}$, has been obtained in [37] by the same method as in [35].

## Astrophysical and cosmological data on the possible variability of the fine structure constant

The question of the possible variability of the constant α and other FFCs in astrophysical conditions involving strong gravitational fields and on cosmological scales of distance and time is widely discussed in the literature [38-45] and has become the subject of many astronomical observations and a thorough analysis of their results. Possible alpha variations are considered in the following astrophysical objects: "ordinary" stars in our and neighboring galaxies [38], white dwarfs [39], distant redshifted galaxies [40], distant redshifted quasars with $z \leq 2,7$ [41], microwave background [42], matter in the process of primary nucleosynthesis [43-45].

It should be noted that the only existing positive result regarding the variability of α was obtained by the group of J. Webb (Australia) [46-48] by analyzing the absorption spectra of various ions in distant quasars. Based on observational data (obtained mostly from the Keck telescope in Hawaii until 2001), it was concluded [46] that the α value billions of years ago was somewhat lower than it is today, with a relative change $\Delta\alpha/\alpha$ in order $10^{-5}$. The data obtained by 2010 from the VLT (Very Large Telescope) located in Chile showed the values of α in the distant past slightly higher than the modern ones. A comparative analysis of the obtained results led the researchers to the conclusion that the parameter α in the past approximately depended on the direction of observations ("the Australian dipole") [47]. Moreover, the axis of the dipole corresponds to the declination $(-61\pm 9)°$ and right ascension of $(17,3\pm 0,6)$ hours with an amplitude of α deviations from its current value $α_0$: $\delta\alpha/\alpha_0$ is of the order of $10^{-6}$ per billion years, which corresponds to the maximum (in direction) change $|\dot\alpha/\alpha|$ of the order of $10^{-15}$/year. The reliability level of this conclusion was estimated by the authors [47] as 4.1σ.

Shortly after the publication of the above results, serious doubts were expressed about their correctness (see, for example, [48]), and their appearance was associated with possible systematic instrumental factors. Nevertheless, the spatial dependence of the α value continues to be discussed up to the present time, and new articles by the Webb group [40, 41, 49, 50] analyze the subtleties of the methodology for studying the emission and absorption spectra of distant galaxies and quasars, confirming and clarifying the previously obtained conclusions about spatial variations of α [6, 50]. For example, in [6] it is stated that "new limits on possible spatial variation of α are preferable to a model without changes of α at the level of 3.7σ".

Other areas of research should also be mentioned. Thus, the search for variations of α in comparison with its laboratory value in the absorption spectra of nearby stars of our galaxy is described in [38], where, as in other studies, the α dependence of spectral line frequencies is used. The influence of systematic effects related to the properties of stellar atmospheres was minimized as a result of the selection of 17 stars with almost identical solar-type atmospheres. The result is a restriction of the form $\Delta\alpha/\alpha \leq 3\times 10^{-8}$ within 50 pc from the Solar System. The article [39] investigated the radiation of white dwarfs, whose strong gravitational fields (with a gravitational potential about 5 orders of magnitude greater than Earth's) could affect the fine structure constant. However, in this case, the conclusions on possible variations of FFCs turn out to be model-dependent, since the ideas about the structure of white dwarfs depend on the choice of the theory of gravity, which was demonstrated in [39] using the examples of general relativity and *f*(*R*) theories. The permissible deviations of α from the laboratory value according to the results of observations of white dwarfs turn out to be within the limits $10^{-4}-10^{-5}$, but with rather large uncertainties due to the indicated model dependence.

The most numerous and informative works are the studies related to the analysis of the spectra of distant galaxies and quasars. In addition to the already mentioned works of the Webb group, let us point out the papers [40, 41], which in general do not contradict the picture of the Australian dipole, but impose restrictions on its magnitude. In particular, according to [40], α variations are within $(1-2) \cdot 10^{-4}$ (1σ) over the last 13.2 billion years. The possible influence of α variability on the cosmic microwave background radiation is considered in some works (see [42] and references therein), however, there, instead of new explicit estimates of such variability, restrictions on the parameters of the theories describing it are discussed, following from the observational data on α variations.

The value of the fine structure constant significantly affects the process of primary nucleosynthesis and its results, such as the number of certain isotopes in the early universe [43, 44]. An analysis of its isotopic composition, in particular, the content of $^2$H and $^4$He, led the authors [43] to conclude that the relative change in α from the time of primary nucleosynthesis to the present does not exceed 2%, which at the time of publication [43] was the most severe constraint. However, other researchers of primary nucleosynthesis come to the conclusion that α values in the corresponding epoch are somewhat lower than in the modern one. Thus, according to [44], the combined latest data on the content of primary deuterium and $^4$He [45] have led to the estimate $-2,6\% < \Delta\alpha/\alpha < -1,4\%$ at a confidence level of 68%. However, it should be mentioned that the α variability is not the only possible explanation for the data on nucleosynthesis; alternative explanations involve assumptions on the lepton asymmetry or a non–standard number of neutrino species [44, 45].

## Conclusion

Currently, there are two most promising methods for determining the fine structure constant, which compete with each other in improving the accuracy of its determination. However, the problem of matching the most highly precision results remains unresolved.

In the widely discussed problem of possible variations of the fine structure constant on a cosmological time scale, the totality of astrophysical data may indicate deviations in the values of α in the distant past of the Universe compared with the modern era, ranging from deviations of the order of one percent in the era of nucleosynthesis to a gradual approach to the modern value.

Despite the lack of high-level experimental evidence on long-term variations of the fine structure constant in laboratory conditions, it can be noted that such stability can be taken into account in practice. For example, it would be promising to create an optical frequency standard with high stability and accuracy of radiation frequency reproduction based on the ytterbium-171 ion and a laser frequency synthesizer. Thus, an optical frequency standard may replace the caesium one.

# References

1. Bureau International des Poids et Measures [official site]. *Resolution 1 of the 26th CGPM (2018). On the revision of the International System of Units (SI).* https://www.bipm.org/en/committees/cg/cgpm/26-2018/resolution1 (date of request: 19.03.2025)

2. Mills I. M., Mohr P. J., Quinn T. J. et al. Redefinition of the kilogram, ampere, kelvin and mole: a proposed approach to implementing CIPM recommendation 1 (CI-2005). *Metrologia*, **43**(3), 227–246 (2006). https://doi.org/10.1088/0026-1394/43/3/006

3. Kononogov S. A. *Metrology and fundamental physical constants*, Standardinform Publ., Moscow (2008) (in Russian).

4. Will C. M. The Confrontation between General Relativity and Experiment. *Living Reviews in Relativity* **9**, 3 (2006). https://doi.org/10.12942/lrr-2006-3

5. Martins C. J. A. P. The status of varying constant: a review of the physics, searches and implications. *Reports on Progress in Physics,* **80**(12), 126902 (2017). https://doi.org/10.1088/1361-6633/aa860e

6. Wilczynska M. R., Webb J. K., Bainbridge M. et al. Four direct measurements of the fine-structure constant 13 billion years ago. *Science Advances*, **6**(17), 9672 (2020). https://doi.org/10.1126/sciadv.aay9672

7. Safronova M. S., Budker D., DeMille D. et al. Search for new physics with atoms and molecules. *Reviews of Modern Physics*, **90**, 025008 (2018). https://doi.org/10.1103/RevModPhys.90.025008

8. Uzan J.-P. Fundamental constants: from measurement to the universe, a window on gravitation and cosmology. *Cosmology and Nongalactic Astrophysics.* https://arxiv.org/abs/2410.07281

9. Sommerfeld A. Zur Quantentheorie der Spektrallinien. *Annalen der Physik*, **366**(51), 1–94 (1916). https://doi.org/10.1002/andp.19163561702

10. Van Dyck R. S., Schwinberg P. B., Dehmelt H. G. New high-precision comparison of electron and positron $g$ factors. *Physical Review Letters*, **59**(1), 26–29 (1987). https://doi.org/10.1103/PhysRevLett.59.26

11. Odom B., Hanneke D., D'Urso B. et al. New measurement of the electron magnetic moment using a one-electron quantum cyclotron. *Physical Review Letters*, **97**(3), 030801 (2006). https://doi.org/10.1103/PhysRevLett.97.030801

12. Gabrielse G., Hanneke D., Kinoshita T. et al. New determination of the fine structure constant from the electron g value and QED (Erratum), *Physical Review Letters*, **99**, 039902 (2007). https://doi.org/10.1103/PhysRevLett.99.039902

13. Hanneke D., Fogwell S., Gabrielse G. New measurement of the electron magnetic moment and the fine structure constant. *Physical Review Letters*, **100**, 120801 (2008).

https://doi.org/10.1103/PhysRevLett.100.120801

14. Fan X., Myers T. G., Sukra B. A. D., Gabrielse G. Measurement of the Electron Magnetic Moment. *Physical Review Letters*, **130**, 071801 (2023). https://doi.org/10.1103/PhysRevLett.130.071801

15. Kinoshita T., Nio M. Improved $\alpha^4$ term of the electron anomalous magnetic moment. *Physical Review D*, **73**, 013003 (2006). https://doi.org/10.1103/PhysRevD.73.013003 ;

16. Aoyama T., Hayakawa M., Kinoshita T. et al. Revised value of the eighth-order electron g-2, *Physical Review Letters*, **99**, 110406 (2007). https://doi.org/10.1103/physrevlett.99.110406

17. Aoyama T., Kinoshita T., Nio M. Revised and improved value of the QED tenth-order electron anomalous magnetic moment. *Physical Review D*, **97**(3), 036001 (2018). https://doi.org/10.1103/PhysRevD.97.036001

18. Wicht A., Hensley J. M., Sarajlic E., Chu S. A preliminary measurement of the fine structure constant based on atom interferometry. *Physica Scripta*, **2002**(T102), 82–88 (2002). https://doi.org/10.1238/Physica.Topical.102a00082

19. Cadoret M., de Mirandes E., Clade P. et al. Combination of Bloch oscillations with a Ramsey-Bordé interferometer: new determination of the fine structure constant. *Physical Review Letters,* **101**, 230801 (2008). https://doi.org/10.1103/PhysRevLett.101.230801

20. Bouchendira R., Cladé P., Guellati-Khélifa S., Nez F., Biraben F. New determination of the fine structure constant and test of the quantum electrodynamics. *Physical Review Letters*, **106**, 080801 (2011). https://doi.org/10.1103/PhysRevLett.106.080801

21. Clade P., de Mirandes E., Cadoret M. et al. Precise measurement of $h/m_{Rb}$ using Bloch oscillations in a vertical optical lattice: determination of the fine-structure constant, *Physical Review A*, **74**, 052109 (2006). https://doi.org/10.1103/PhysRevA.74.052109 ;

22. Parker R. H., Yu C., Zhong W., Estey B., Müller H. Measurement of the fine-structure constant as a test of the Standard Model. *Science*, **360**(6385), 191–195 (2018). https://doi.org/10.1126/science.aap7706

23. Morel L., Yao Z., Cladé P., Guellati-Khélifa S. Determination of the fine-structure constant with an accuracy of 81 parts per trillion. *Nature*, **588**, 61–65 (2020). https://doi.org/10.1038/s41586-020-2964-7

24. Borde Ch. J. Atomic interferometry with internal state labeling. *Physics Letters A*, **140**(1-2), 10–12 (1989). https://doi.org/10.1016/0375-9601(89)90537-9

25. Tiesinga E., Mohr P. J., Newell D. B., Taylor B. N. CODATA recommended values of the fundamental physical constants: 2018. *Reviews of Modern Physics*, **93**, 025010 (2021). https://doi.org/10.1103/RevModPhys.93.025010

26. Mount B. J., Redshaw M., Myers E. G. Atomic masses of $^6$Li , $^{23}$Na, $^{39,41}$K, $^{85,87}$Rb, and $^{133}$Cs. *Physical Review A*, **82**, 042513 (2010). https://doi.org/10.1103/PhysRevA.82.042513

27. Mohr P. J., Newell D. B., Taylor B. N. CODATA recommended values of the fundamental physical constants: 2014. *Reviews of Modern Physics*, **88**, 035009 (2016).

https://doi.org/10.1103/RevModPhys.88.035009

28. Tanabashi M., Hagiwara K., Hikasa K. et al., Review in Particle Physics. *Physical Review D*, **98**, 030001 (2018). https://doi.org/10.1103/PhysRevD.98.030001

29. Schwinger J. On Quantum-Electrodynamics and the Magnetic Moment of the Electron. *Physical Review Journals Archive*, **73**, 416 (1948). https://doi.org/10.1103/PhysRev.73.416

30. Bronnikov K. A., Ivashchuk V. D., Khruschov V. V. Fundamental physical constants: search results and descriptions of variations. *Measurement Techniques*, **65**(3), 151–156 (2022). https://doi.org/10.1007/s11018-022-02062-z

31. Bronnikov K. A., Kalinin M. I., Khruschov V. V. On the thermal history of the early Universe. *Legal & Applied Metrology*, (1), 11–17 (2024) (in Russian); arXiv: 2312.12883.

32. Rosenband T., Hume D. B., Schmidt P. O. et al. Frequency Ratio of Al+ and Hg+ Single-Ion Optical Clocks; Metrology at the 17th Decimal Place. *Science*, **319**(5871), 1808–1812 (2008). https://doi.org/10.1126/science.1154622

33. Godun R. M., Nisbet-Jones P. B. R., Jones J. M. et al. Frequency ratio of two optical clock transitions in $^{171}$Yb$^+$ and constraints on the time variation of fundamental constants. *Physical Review Letters*, **113**, 210801 (2014). https://doi.org/10.1103/PhysRevLett.113.210801 ;

34. Levshakov S. A., Ng K-W., Henkel C. et al. Testing the weak equivalence principle by differential measurements of fundamental constants in the Magellanic Clouds. *Monthly Notices of the Royal Astronomical Society*, **487**(4), 5175–5187 (2019). https://doi.org/10.1093/mnras/stz1628

35. Lange R., Huntemann N., Rahm J. M. et al. Improved Limits for Violations of Local Position Invariance from Atomic Clock Comparisons. *Physical Review Letters,* **126**, 011102 (2021). https://doi.org/10.1103/physrevlett.126.011102

36. Flambaum V. V., Dzuba V. A. Search for variation of the fundamental constants in atomic, molecular and nuclear spectra. *Canadian Journal of Physics.*, **87** (1), 25–33 (2009). https://doi.org/10.1139/p08-072

37. Filzinger M., Dorscher S., Lange R. et al. Improved Limits on the Coupling of Ultralight Bosonic Dark Matter to Photons from Optical Atomic Clock Comparisons. *Physical Review Letters,* **130**, 2530011 (2023). https://doi.org/10.1103/PhysRevLett.130.253001

38. Murphy M. T., Berke D.A., Liu F. et al. A limit on variations in the fine-structure constant from spectra of nearby Sun-like stars. *Science*, **378** (6620), 634–636 (2022). https://doi.org/10.1126/science.abi9232

39. Kalita S., Uniyal A. Constraining fundamental parameters in modified gravity using Gaia-DR2 massive white dwarf observation. *The Astrophysical Journal,* **949**(2), 62 (2023). https://doi.org/10.3847/1538-4357/accf1c

40. Jiang L., Fu S., Wang F. et al. Constraints on the variation of the fine-structure constant at 3 < z < 10 with JWST emission-line galaxies. *Cosmology and Nongalactic Astrophysics* (2024). https://arxiv.org/abs/2405.08977

41. Milakovic D. Fine structure constant measurements in quasar absorption systems.

*Methodology* (2023). https://arxiv.org/abs/2310.01071

42. Tohfa H., Crump J., Baker E. et al. A cosmic microwave background search for fine-structure constant evolution. *Cosmology and Nongalactic Astrophysics* (2023). https://arxiv.org/abs/2307.06768

43. Meisner U.-G., Metsch B. Ch., Meyer H. The electromagnetic fine-structure constant in primordial nucleosynthesis revisited. *High Energy Physics – Theory* (2023). https://arxiv.org/abs/2305.15849

44. Seto O., Takahashi T., Toda Y. Variation of the fine structure constant in the light of recent helium abundance measurement. *Physical Review D*, **108**, 023525 (2023). https://doi.org/10.1103/PhysRevD.108.023525

45. Matsumoto A., Ouchi M., Nakajima K. et al., EMPRESS. VIII. A new determination of primordial He abundance with extremely metal-poor galaxies: a suggestion of the lepton asymmetry and implications for the Hubble tension. *The Astrophysical Journal*, **941**(2), 167 (2022). https://doi.org/10.3847/1538-4357/ac9ea1

46. Webb J. K., Murphy M. T., Flambaum V. V. et al. Further evidence for cosmological evolution of the fine structure constant. *Physical Review Letters*, **87**, 091301 (2001). https://doi.org/10.1103/PhysRevLett.87.091301

47. Webb J. K., King J.A., Murphy M. T. et al. Indications of a Spatial Variation of the Fine Structure Constant. *Physical Review Letters*, **107**, 191101 (2011). https://doi.org/10.1103/PhysRevLett.107.191101

48. Levshakov S. A., Combes F., Boone F. et al., An upper limit to the variation in the fundamental constants at redshift z=5.2. *Astronomy and Astrophysics,* **540**, L9 (2012). https://doi.org/10.1051/0004-6361/201219042

49. Whitmore J. B., Murphy M. T., Impact of instrumental systematic errors on fine-structure constant measurements with quasar spectra. *Monthly Notices of the Royal Astronomical Society*, **447**(1), 446–462 (2015). https://doi.org/10.1093/mnras/stu2420

50. Lee C.-C., Webb J. K., Milaković D., Carswell R. F. Non-uniqueness in quasar absorption models and implications for measurements of the fine structure constant, *Monthly Notices of the Royal Astronomical Society*, **507** (1), 27–42 (2021). https://doi.org/10.1093/mnras/stab2005